%% file: main.tex
\documentclass{article}
\usepackage{arxiv}
\usepackage[utf8]{inputenc} 
\usepackage[T1]{fontenc}    
\usepackage{url}            
\usepackage{booktabs}       
\usepackage{amsfonts}       
\usepackage{nicefrac}       
\usepackage{microtype}      
\usepackage{lipsum}	
\usepackage{float}
\usepackage{doi}
\usepackage{xcolor}
\usepackage{authblk}
\usepackage{comment}
\usepackage{float}
\usepackage{listings}
\usepackage{caption}
\usepackage{subcaption}
\usepackage{graphicx}

\definecolor{codegreen}{rgb}{0,0.6,0}
\definecolor{codegray}{rgb}{0.5,0.5,0.5}
\definecolor{codepurple}{rgb}{0.58,0,0.82}
\definecolor{backcolour}{rgb}{0.95,0.95,0.92}
 
\lstdefinestyle{mystyle}{
    backgroundcolor=\color{backcolour},   
    commentstyle=\color{codegreen},
    keywordstyle=\color{magenta},
    numberstyle=\tiny\color{codegray},
    stringstyle=\color{codepurple},
    basicstyle=\footnotesize,
    breakatwhitespace=false,         
    breaklines=true,                 
    captionpos=b,                    
    keepspaces=true,                                  
    numbersep=5pt,                  
    showspaces=false,                
    showstringspaces=false,
    showtabs=false,                  
    tabsize=2
}

\title{Generative AI Toolkit - a framework for increasing the quality of LLM-based applications over their whole life cycle}

\date{}

\author[1]{Jens Kohl} 
\author[1]{Luisa Gloger}
\author[2]{Rui Costa} 
\author[2]{Otto Kruse} 
\author[1]{Manuel P. Luitz}
\author[1]{David Katz}
\author[2]{Gonzalo Barbeito}
\author[2]{Markus Schweier}
\author[2]{Ryan French}
\author[1]{Jonas Schroeder}
\author[1]{Thomas Riedl}
\author[1]{Raphael Perri}
\author[1]{Youssef Mostafa}

\affil[1]{BMW Group, Munich, Germany}
\affil[2]{Amazon Web Services}

\hypersetup{colorlinks=true, urlcolor=blue, linkcolor=black}

\begin{document}

\lstset{language=Python}

\maketitle

\begin{abstract}

As LLM-based applications reach millions of customers, ensuring their scalability and continuous quality improvement is critical for success. However, the current workflows for developing, maintaining, and operating (DevOps) these applications are predominantly manual, slow, and based on trial-and-error. With this paper we introduce the Generative AI Toolkit, which automates essential workflows over the whole life cycle of LLM-based applications. The toolkit helps to configure, test, continuously monitor and optimize Generative AI applications such as agents, thus significantly improving quality while shortening release cycles. We showcase the effectiveness of our toolkit on representative use cases, share best practices, and outline future enhancements. Since we are convinced that our Generative AI Toolkit is helpful for other teams, we are open sourcing it on  \href{https://github.com/awslabs/generative-ai-toolkit}{GitHub} and hope that others will use, forward, adapt and improve.

\end{abstract}

\keywords{LLM-based Applications \and LLM-based agents \and LLM-DevOps Framework \and LLM-Application Framework}

\input{intro}
\input{relatedwork}
\input{contribution}
\input{eval}

\input{conclusion}

\bibliographystyle{ieeetr} 
\bibliography{main} 
\input{appendix}

\end{document}

%% file: intro.tex
\section{Introduction} \label{sec:intro}

\subsection{Challenges in developing and operating LLM-based applications over their whole life cycle}

Large Language Models (LLM) are machine learning models capable of executing natural language processing tasks such as summarizing or generating text. Since their introduction LLM have gained widespread traction in different domains. They can be used as stand-alone products, but also to augment existing software products such as applications (also called agentic functions) or machine learning agents (also called LLM-based agents) to increase their capabilities.  

In this section, we show challenges during development and operation of LLM-based applications on three examples.

Users interact with LLM-based applications by entering input into the LLM, the so-called prompt. Jang et al. showed in 2023 that the LLM’s output is very sensitive to variations of the prompt \cite{jang2023can}. Thus, the task of finding the best prompt to generate expected or best output leads to manual, trial-and-error-prompt experimenting – a method well known as prompt-engineering (cf. White et al. in 2023 for ChatGPT \cite{white2023promptpatterncatalogenhance} or a survey of prompt techniques by Schulhoff et al. in 2024 \cite{schulhoff2024promptreportsystematicsurvey}).

Additionally, the outputs of an LLM-based application can not only vary, but also be wrong without telling a user (“hallucination”, cf. Sec 2.). This means that LLM-based applications need to be tested sufficiently during development which is aggravated due to limited insight into the LLM-based application. Keeping track of various prompt versions during prompt engineering and comparing results over a large test dataset adds even more development complexity.

Given the fact that LLM-based applications can also interact autonomously with potentially millions of users, they need to be monitored at scale after deployment and release to prevent hallucination and identify other quality-related issues.

In traditional software development similar challenges are addressed by tool-supported Continuous Integration / Continuous Deployment (CI/CD) pipelines with high degree of automation. In contrast, given the nascent state of LLM-based applications, automated tooling or workflows for their development and operation are either premature products or nonexistent. Moreover, as most of the existing tooling focuses on selected aspects of the application life cycle, comprehensive frameworks covering the whole life cycle end-to-end are scarce (cf. Sec. \ref{sec:relatedwork} for a detailed review).

\subsection{Overview of our contribution}

In this paper we introduce a toolkit automating DevOps workflows for LLM-based applications with the goal of increasing the quality of LLM-based applications over their whole life cycle while introducing more efficient workflows. We show the benefits of the Generative AI Toolkit on representative use cases. With these use cases we also share some best practices for optimizing LLM-based applications over their whole life cycle. Since our use cases are using single LLM or LLM-based agents, we cover the currently most common architectural patterns for LLM-based applications. 

%% file: relatedwork.tex
\section{Related Work} \label{sec:relatedwork}

In this section we give some background regarding relevant software engineering workflows and then detail LLM-based agents and their history. Furthermore, we show that there are currently few open source and free of charge frameworks covering the whole LLM-agent life cycle.

\paragraph{DevOps:} is an approach aiming to solve the previous separation between engineers developing software and engineers operating it. The idea was that engineers should own their software and thus ensure quality over the whole life cycle. One of the first mentions of this term was in a conference in Gent in 2009. An example for state of the art and challenges for DevOps can be found in Leite et al. in 2019 \cite{leite2019survey}. These techniques have been adopted to operate machine learning systems as we will show in a later paragraph of this section.

\paragraph{Continuous Integration and Continuous Delivery (CI/CD):} Continuous Integration, as defined by Fowler in 2006, is a practice “where members of a team integrate their work frequently...to multiple integrations a day” leading to “significantly reduced integration problems and allows to develop cohesive software more rapidly \cite{fowler2006continuous}”. Beck et al. defined in 2001 Continuous Deployment as the next step with the goal to “satisfy the customer through early and continuous delivery of valuable software” \cite {beck2001agile} by a “reliable...and largely automated process” as Humble and Farley stated in 2010 \cite{humble2010continuous}). Both approaches combined offer reduced time to market and an improved efficiency and quality of the software (e.g. Chen in 2015 \cite{chen20215continuous}). Core enabler of CI/CD is a pipeline in which all essential workflows for building, integrating, deploying, and releasing the software are fully automated. 

\paragraph{Software testing:} is one of the essential steps of the CI/CD pipeline. Software testing has been extensively researched with a substantial body of literature addressing efficient software testing, e.g., Dustin et al. in 1999 \cite{dustin1999automated} or Gregory and Crispin in 2014 \cite{gregory2014more}. A main concept is the test pyramid with at least three layers as introduced by Kohn in 2010 starting with single tests of small software functions (unit tests), followed by testing of the interactions of these small functions (integration tests) and ending with testing the whole system (acceptance tests) \cite[Chap. 16]{cohn2010succeeding}. This test cascade is nowadays executed automatically as part of a CI/CD pipeline. Since only a successful passing of all tests of one layer leads to testing the next layers, this concept helps to provide a good trade-off between testing time and test depth (although depending on the quality of the tests). 

\paragraph{Large-language models (LLM):} we show a brief overview of selected LLM and focus on the improvement of model quality; a very comprehensive overview for LLM can be found under Zhao et al. from 2023 \cite[Sec 2.2]{zhao2023survey}.

Most LLM use Transformers with attention layers as the basis of their architecture. Attention layers were first described by Schmidhuber in 1992 \cite{schmidhuber1992learning} and by Bahdanau et al. in 2015 \cite{bahdanau2015}. In 2017, Vaswani et al. refined the attention layer and built the structure now known as Transfomer \cite{vaswani2017attention}. In contrast to previously existing architectures such as long short-term memory \cite{hochreiter1997long}, attention layers allow for more efficiency than previous architectures in training since compute load can be distributed on multiple machines.

OpenAI’s publication of GPT1 by Radford et al. in 2018 made LLM quite popular \cite{radford2018improving}. OpenAI published subsequent papers for GPT2 (Radford et al. in 2019 \cite{radford2019language}) and GPT3 (Brown et al. in 2020 \cite{brown2020language}) increasing the capabilities of the model and its qualitative output by using more training data and more parameters. In 2022, Open AI’s Instruct GPT introduced reinforcement learning to improve the quality of model output (Ouyang et al. \cite{ouyang2022training}).  

Meta's open sourcing of development steps, source code and weights of Llama 1 and Llama 2 in 2023 (Touvron et al. \cite{touvron2023llama1, touvron2023llama2}) were a big step to accelerate LLM development. Since then, several papers have been published improving the quality of the model and its efficiency. Deep Mind's Chinchilla \cite{hoffmann2022training} established a mathematical relation between the number of tokens a model was trained on, its parameter size and the quality of its outputs. Based on this relation, several papers showed that models with smaller parameter size can have comparably good quality, e.g., Phi by Gunsekar et al. \cite{gunasekar2023textbooks}. Through the Falcon model, Penedo et al. \cite{penedo2023refinedweb} showed the impact of curated data on a model's quality.

\paragraph{LLM-based agents:} Russel and Norvig \cite[Chapter 2]{russell-norvig2021} as well as Xi et al. in 2023 \cite[Sec. 2]{xi2023rise} show an overview of the history of agents in machine learning. 

LLM-based agents extend machine learning agents by integrating a LLM to "...effectively perform diverse tasks by leveraging the human-like capabilities of [a] LLM" (Wang et al. 2024, \cite{wang2024survey}). 
Weng (2023) gives a graphical overview and description of an agent and its components. An agent can access, activate, or interact with these other agent related components to resolve the requests given by a user's textual input, the \textbf{prompt}. The agent uses \textbf{planning} to break down its task(s) given via prompt into steps or subtasks. That can be \textbf{reflection} or \textbf{self-criticism} to increase the quality of the final output of the agent by learning from history. For this, the agent can use \textbf{short-term memory} (e.g. chat history) but can also access \textbf{long-term memory}, i.e., external knowledge stored in databases. \textbf{Tools} are external API functions the agent can call to get external information or execute code \cite{weng2023llm}. 

Overviews of current and historical LLM-based agents can be found in Xi et al. and Naveed et al. (both in 2023, \cite{xi2023rise, naveed2023comprehensive}),  or Wang et al. in 2024 \cite{wang2024survey}. Exemplary frameworks are Wang et al. (2023) \cite{wang2023describe}, Yao et al. (2023) \cite{yao2023reactsynergizingreasoningacting} in which LLM are used for planning or Wu et al. (2023) \cite{wu2023autogen} for multi-agents. 

\paragraph{Metrics for LLM-based applications:} most of the existing metrics focus on the evaluation of the output of an agent. Well-known examples for natural-language relevant metrics are Bilingual Evaluation Understudy (BLEU) \cite{papineni2002bleu} measuring similarity between a generated and reference sentence, Recall Oriented Understudy for Gisting Evaluation (ROUGE) \cite{lin2005recall} for text summarization or General Language Understanding Evaluation (GLUE) \cite{wang2018glue}. Banerjee et al. (2023) \cite{banerjee2023benchmarking} and Bandi et al. (2023) \cite{bandi2023power} give overviews of metrics for LLM-powered chatbots, Wandi (2022) \cite{wang2022modern} shows existing datasets for evaluations.

\paragraph{Testing LLM-based applications:} as LLM-based applications, and especially agents,  can interact and support humans directly without supervision, they need to be tested intensively for characteristic quality problems.

A very well-known problem is “hallucination”, i.e. "the generation of texts or responses that exhibit grammatical correctness, fluency, and authenticity, but deviate from the provided source inputs (faithfulness) or do not align with factual accuracy (factualness)" (Ji et al. (2023) \cite{ji2023survey}). Although several techniques have empirically shown to reduce hallucination, e.g. the popular use of Retrieval Augmented Generation (RAG) to dynamically contextualize an LLM response from a knowledge base as shown by Shuster in 2021 \cite{shuster2021retrieval}, Xu et al. and Banerjee et al. both showed in 2024 that hallucination is impossible to avoid completely \cite{xu2024hallucinationinevitableinnatelimitation, banerjee2024llmshallucinateneedlive}. 

Another example is “jailbreaking”, i.e., circumventing measures to prevent undesirable generation \cite{zou2023universal}. Jailbreaking was first shown by Zou et al. in 2023 \cite{zou2023universal}, Xu et al. (2024) \cite{xu2024comprehensivestudyjailbreakattack} gives an overview of possible jailbreaks. Although most jailbreak attempts focus on prompts to the LLM, Deng et al. (2024) showed that this is also possible via RAG \cite{deng2024pandorajailbreakgptsretrieval}.

Lastly, LLM's output and thus accuracy was shown by Jang et al. in 2023 to be very sensitive to prompt variation \cite{jang2023can}. 

Currently, most research on testing agents (e.g., Chang et al. or Aleti et al. both in 2023  \cite{chang2023surveyevaluationlargelanguage, aleti2023softwaretestinggenerativeai}) or frameworks for this (e.g. \href{https://docs.confident-ai.com/}{DeepEval} or \href{https://docs.ragas.io/en/stable/}{Ragas}) focuses on acceptance level; that is entering an input or prompt into the agent and comparing the output with expected output. This allows testing from the perspective of a future user, but gives only limited insight into the agent’s components and interactions leading to a possible unexpected behavior. Additionally, acceptance tests are often time-consuming and difficult to automate.

\paragraph{Machine Learning and LLM operations:} Machine Learning Operations (MLOps) represents a systematic approach to industrializing ML systems through standardized processes and automated workflows, introduced with the goal of bridging the gap between model development and operational deployment. At its core, this discipline combines software engineering best practices with ML-specific requirements, establishing frameworks for Continuous Integration/ Deployment, and monitoring of ML models while simultaneously addressing unique challenges such as experiment tracking, model registry management, and automated retraining of pipelines. More in-depth explanations are provided by Renggli et al. in 2019 \cite{renggli2019continuousintegrationmachinelearning} and Kreuzberger et al. in 2023 \cite{kreuzberger2023machine}. Sculley et al. established the foundational challenges of operationalizing Machine Learning systems in 2015 \cite{sculley2015hidden}.

In contrast, few publications have detailed the operations of LLM or the management of their entire life cycle. In fact, Diaz et al. (2024) state there is no common scope for the term LLMOps \cite{diaz2024large}. Zhao et al. explain in their 2023 paper the different life cycle stages without explicitly defining a term or scope \cite{zhao2023survey}.
 
Even though extensive research to improve model quality has been done, the varying output of LLM introduce new complexities to control, test and explain their results pushing the boundaries of classical MLOps techniques (see Bommasani et al. in 2021 \cite{bommasani2021opportunities}, Aleti and Liu et al. in 2023 \cite{aleti2023softwaretestinggenerativeai, liu2023pre} for more context and a comprehensive analysis).

\paragraph{LLM-frameworks:} Generative AI and LLM in particular are nascent technologies which also applies to methods and frameworks for their development and operation. Though academia, industry and communities are moving quickly with lots of development for tooling and frameworks. As of the time of writing this paper, the existing tooling can be categorized in three main groups:

\begin{itemize}
    \item Tools from cloud service providers, such as Amazon Bedrock Model Evaluation or Azure ML Prompt Flows, integrated into their machine learning offerings or Nvidia’s \href{https://github.com/NVIDIA/NeMo}{NVIDIA Nemo Evaluator}. These tools are tailored and optimized for respective stacks.
    \item Commercial off-the-shelf software focused on specific workflows (e.g. \href{https://www.deepchecks.com/}{DeepChecks} for evaluation) or even up to the whole life cycle of an LLM (e.g. \href{https://www.confident-ai.com/}{Confident AI}, \href{https://www.langchain.com/langsmith}{LangSmith}, \href{https://wandb.ai/site/}{Weights and Biases}). These offerings are not (completely) open source (though often based on open source projects) and not free of charge.
    \item Open source: several frameworks are open source and -similar to the commercial offerings- cover some workflows (e.g. \href{https://docs.confident-ai.com/}{DeepEval} or \href{https://docs.ragas.io/en/stable/}{Ragas} for evaluation, DSpy for prompting \cite{khattab2023dspy}) or even the whole life cycle (e.g. \href{https://www.comet.com/site/products/opik/}{Opik}, \href{https://mlflow.org/}{MLFlow} or our toolkit).  
\end{itemize}

At the time of writing this paper, the Generative AI Toolkit covers the whole life cycle similar to Opik, MLFlow, LangSmith, Weights and Biases. Of these frameworks Opik, MLFlow and Generative AI Toolkit are open source; and only the Generative AI Toolkit and MLFlow are free of charge.

%% file: contribution.tex
\section{Contribution} \label{sec:contribution}

\subsection{Overview of our contributions}

\begin{figure}[ht]
    \centering
    \includegraphics[width=0.8\textwidth]{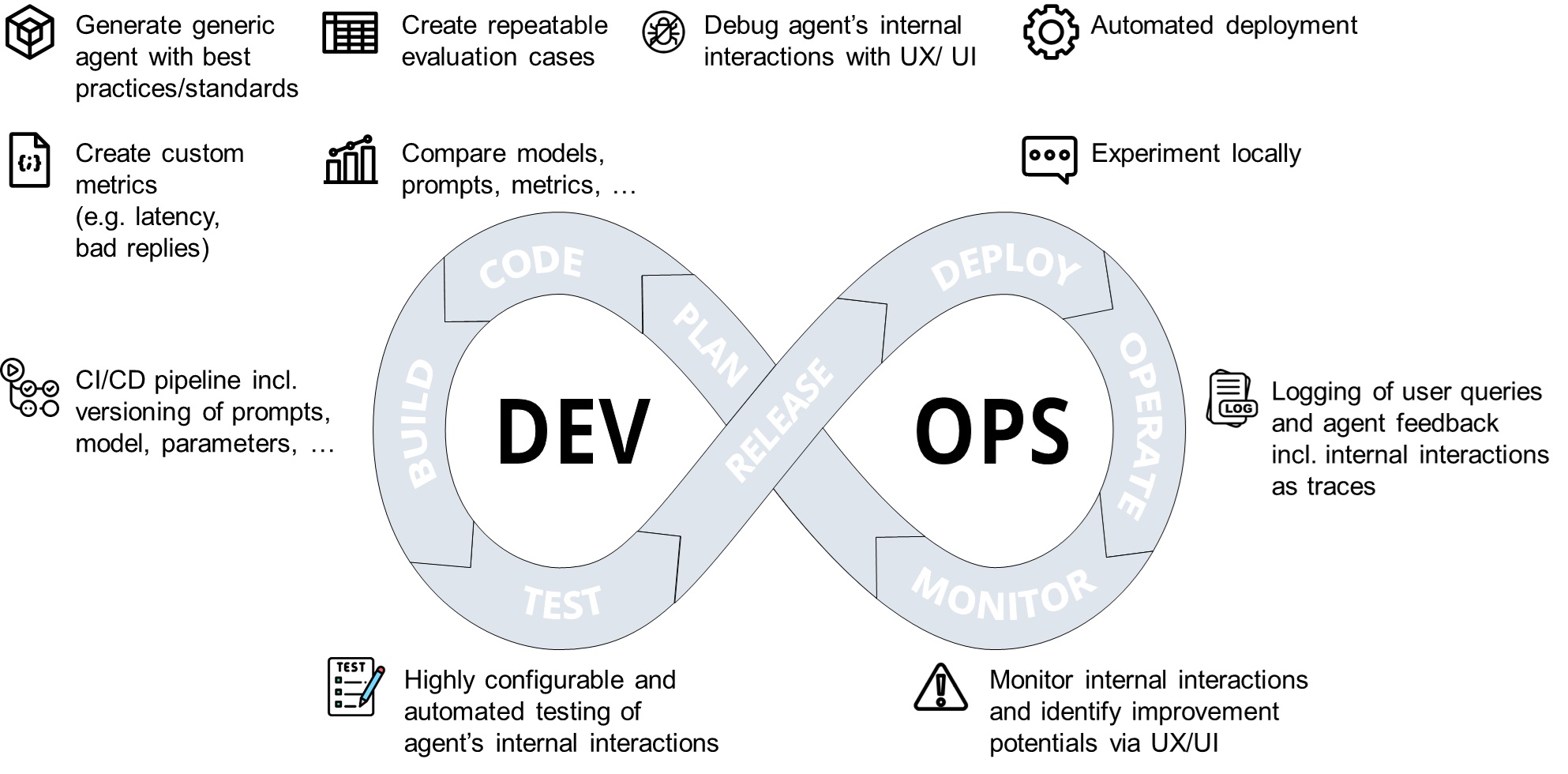}
    \caption{Overview Generative AI Toolkit and its features.}
    \label{fig:GenAI-toolkit-features}
\end{figure}

Figure \ref{fig:GenAI-toolkit-features} shows an overview of the features of the proposed Generative AI Toolkit. It covers the whole DevOps life cycle of a LLM-based application, but focuses on the phases code, build, test, and monitoring.

\subsection{Walkthrough of Generative AI toolkit’s features}

\paragraph{Automate agent creation:} we built a template for LLM-based agents which can be created and deployed with the Python library \href{https://www.cookiecutter.io/}{Cookiecutter} to bootstrap new projects for agent development. The included “vanilla agent” is based on best practices and contains sample code, notebooks, and tools, making it easy for developers to experiment with different prompts, tooling, or response patterns. Furthermore, it allows to run the agent against different metrics (e.g. latency) as well as to deploy and test changes rapidly. 

\paragraph{Define and create custom, user-defined metrics:} agents often require highly customized evaluation methods tailored to their specific goal and intended way of working. The toolkit allows developers to create user-defined metrics that can assess and evaluate the agent as well as its interaction with a human by measuring various factors, such as: the agent's response, tool usage, number of conversation turns or "hops” in the interaction between agent and user, input and output token counts, per-turn costs, latency (total time taken for agent interaction), etc. User-defined metrics can be simple ones such as counting tokens or complex ones such as using an additional LLM as a judge to score the generated output (as proposed by Zheng et al. in 2023 \cite{zheng2023judging}). Since the metrics can be evaluated and analyzed in the development as well as operation phase, they offer a feedback loop to continuously improve the agent’s performance and quality.

\paragraph{Create repeatable evaluation cases:} the feature allows users to conduct comprehensive and standardized evaluations by defining multiple test cases. This includes testing various models, user input scenarios, and customized prompts. These evaluations help ensure consistency across agent versions, making it easier to identify unintended behavioral changes over time. By supporting systematic testing, this feature enables teams to verify that each version meets performance standards and aligns with the intended user experience.

\paragraph{Compare different models and prompts against evaluation cases and metrics:} a challenge during development of LLM-based agents is to find the "best" set of agent parameters. For example, tweaking the agent's system prompt, or one of its tool descriptions, may make some use cases better, but degrade others, so rigorous testing must consider all use cases. Our toolkit therefore allows developers to "permute" agent parameters, and run test cases against each possible permutation, by creating an agent instantiation for each permutation running as a thread. Developers can thus test different language models against a set of slightly different system prompts to find the best prompt-model combination. Likewise other agent parameters can be "permuted" too, such as the temperature parameter, the set of tools available, the top-p parameter, etc. The toolkit will run the developer-defined test cases against each permutation, in parallel against each user-defined metric, and returns a Pandas DataFrame with all measurements, grouped per parameter permutation. This helps developers to identify the best set of parameters.

\paragraph{Tests:} as mentioned in Sec. \ref{sec:relatedwork}, most testing of LLM-based agents is done on acceptance level and in manual workflows or tool-supported (e.g. \href{https://docs.confident-ai.com/}{DeepEval} or \href{https://docs.ragas.io/en/stable/}{Ragas}) against standard benchmarks. We combine the Generative AI toolkit with a new approach for testing LLM-based agents such as the test automation pyramid from software engineering. While this study only provides an introduction and short overview of the framework, more details will be provided in a subsequent publication. The idea of the approach is to first test components of the agent in isolation in automated tests and afterwards test the interactions of these components. With the Generative AI Toolkit, it is possible to run tests of agents against typical benchmarks such as GLUE or BLEU (cf. Sec. \ref{sec:relatedwork}), user-defined evaluation cases and metrics, as well as isolated tests of an agent’s components and their interactions automatically, which are then verified via the traces (see below in paragraph Logging and Monitoring).

\paragraph{Deploy:} the Generative AI toolkit offers Infrastructure-as-Code based on the AWS Cloud Development Kit (\href{https://aws.amazon.com/de/cdk/}{AWS CDK}). By using Cookiecutter, a GitHub workflow YAML file for a simple CI/CD pipeline is created, which can serve as a starting point for a more elaborate pipeline. 

\begin{figure}[!htb]
    \centering
    \includegraphics[width=0.7\textwidth]{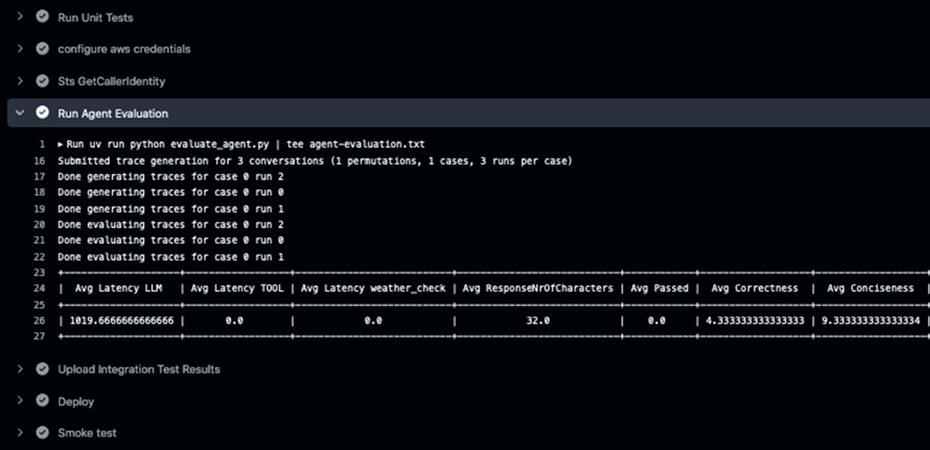}
    \caption{Screenshot of Generative AI Toolkit's CI/CD pipeline.}
    \label{fig:GenAI-toolkit-CICD}
\end{figure}

\paragraph{CI/CD:} the toolkit's evaluation can also be run as part of continuous integration scripts and CI/CD pipelines. Test cases defined by developers can, like unit tests, be run against the agent's implementation in the commit under test. Since our agent's implementation can run locally, cloud deployment of the agent (again, like unit tests) is not necessarily required, except from online access to the models. The resulting traces can then be evaluated against all defined metrics. Assertions can be made on the resulting measurements, so that the commit under test can be properly evaluated, e.g., by comparing measurements automatically against set thresholds, or simply as further information for human reviewers of the associated pull request. It is of course helpful for human reviewers to be able to compare the agent's measurements of the commit in the pull request against the measurements of the latest commit from the target branch. Note that such comparisons are often not Pareto-optimal: some cases may actually achieve worse measurements, but that may be offset by higher measurements on other cases. The toolkit exposes the full set of measurements as a Pandas DataFrame for developers to consume, inspect, and write assertions against. An out-of-the-box summary contains the averages for all custom metrics that were included in the evaluation and can be logged as a table to standard out (cf. Fig. \ref{fig:GenAI-toolkit-CICD}).

\clearpage

\begin{figure}[h]
    \begin{subfigure}{0.5\textwidth}
        \includegraphics[width=0.9\linewidth, height=6cm]{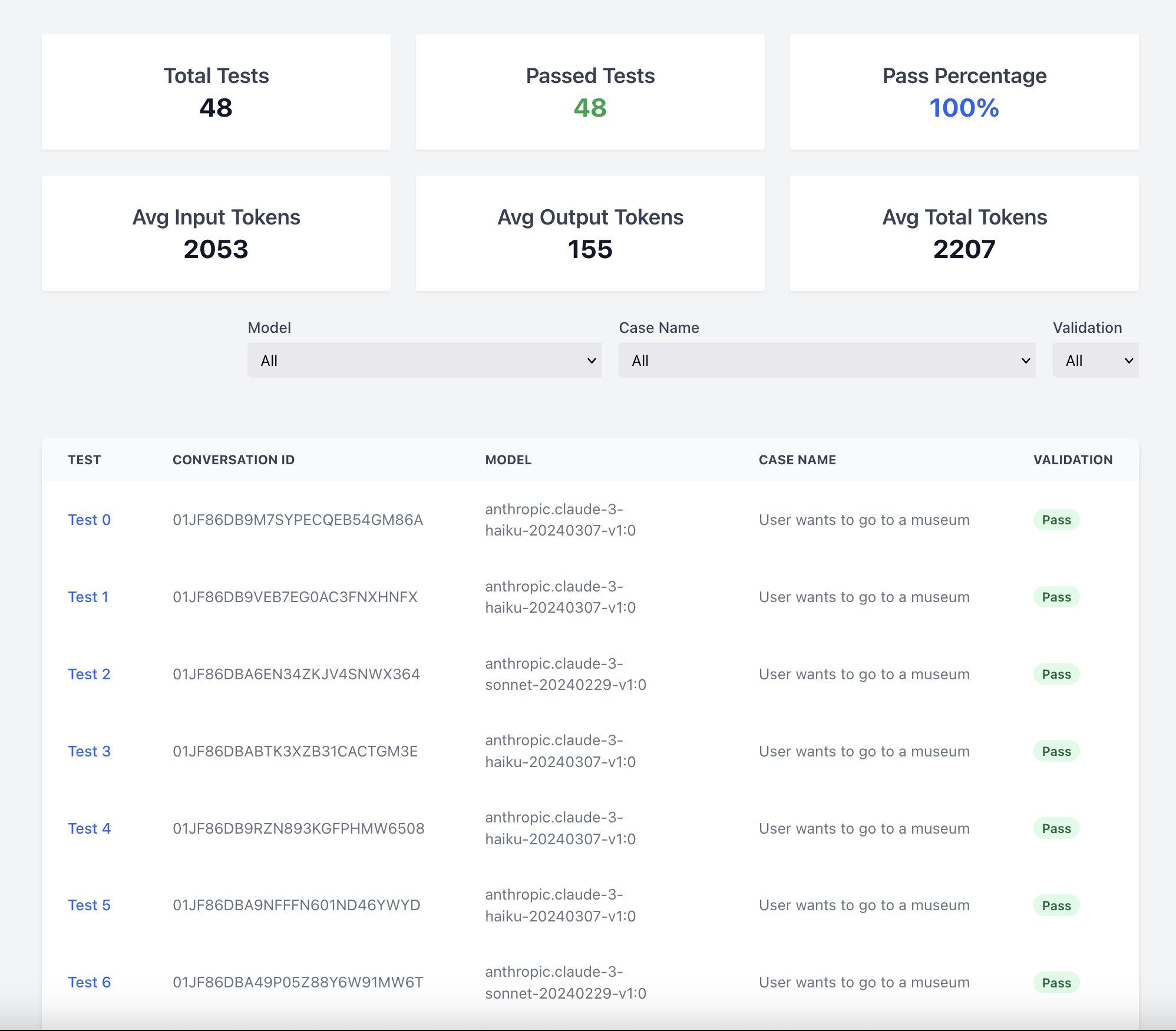} 
        \caption{Screenshot overview of test \& evaluation cases and results.}
        \label{fig:subim1}
    \end{subfigure}
    \hfill
    \begin{subfigure}{0.5\textwidth}
        \includegraphics[width=0.9\linewidth, height=6cm]{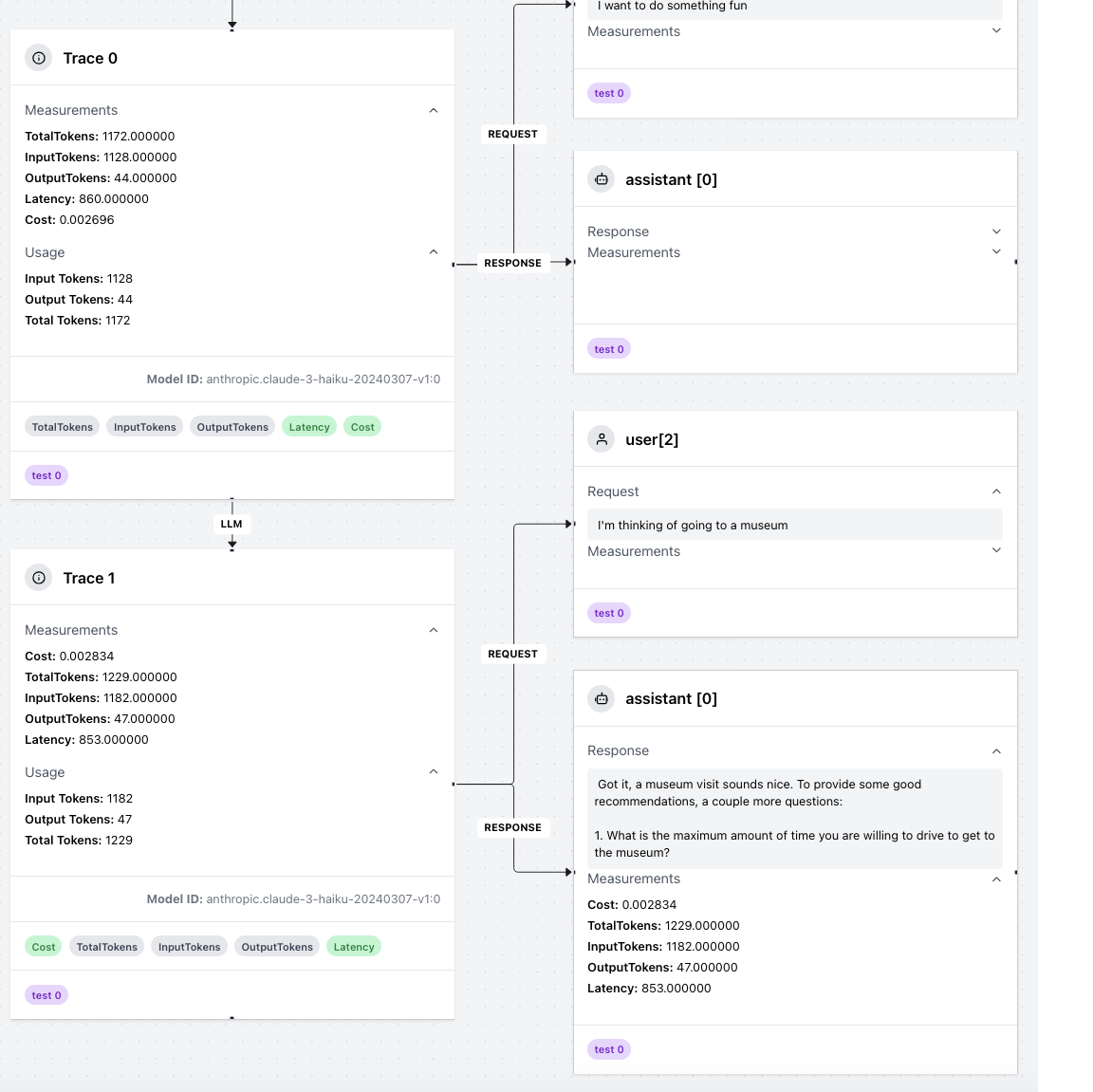}
        \caption{Screenshot of traces, logs and metrics for one test case.}
        \label{fig:subim2}
    \end{subfigure}
    \hfill
\caption{Screenshot of Generative AI Toolkit’s GUI for "debugging" agents.}
\label{fig:GenAI-toolkit-Debugging}
\end{figure}

\paragraph{Graphical User Interface (GUI) for “debugging” the agent:} the GUI of the Generative AI Toolkit is designed to set up a local web server using \href{https://gunicorn.org/}{Gunicorn}, giving developers a clear and interactive view of all evaluations and traces. This GUI allows users to inspect detailed trace logs, verify evaluation results, and quickly check the pass/fail status of tests (cf. Fig. \ref{fig:GenAI-toolkit-Debugging}). By providing a real-time display of test outcomes, the GUI enhances the debugging and quality 
assurance process, making assessing an agent's readiness for deployment easier.

\paragraph{Logging and Monitoring at scale via traces:} the toolkit includes a wrapper around the agent and thus captures detailed logs and metrics for each interaction inside an agent as well as throughout each stage of an agent’s life cycle. The logs are currently stored in DynamoDB, but an extension to other services for observing and storing measurements is easily possible.

This feature allows users to monitor key metrics, track agent interactions, assess response quality, and analyze latency, including end-to-end latency for tool execution. Tracing supports robust debugging and optimization, providing teams with a fine-grained view of how the agent performs under various conditions and helping them identify potential areas for improvement. Typical optimization potentials are when the agent does not have an answer to a request. This can be detected either via checking and triggering for words such as “unfortunately” or “sorry” (cf. Appendix \ref{subsec:appendix:use-cases:uc3}) or by checking if the same user asks similar queries in a short time. 

\begin{figure}[!htb]
    \centering
    \includegraphics[width=0.7\textwidth]{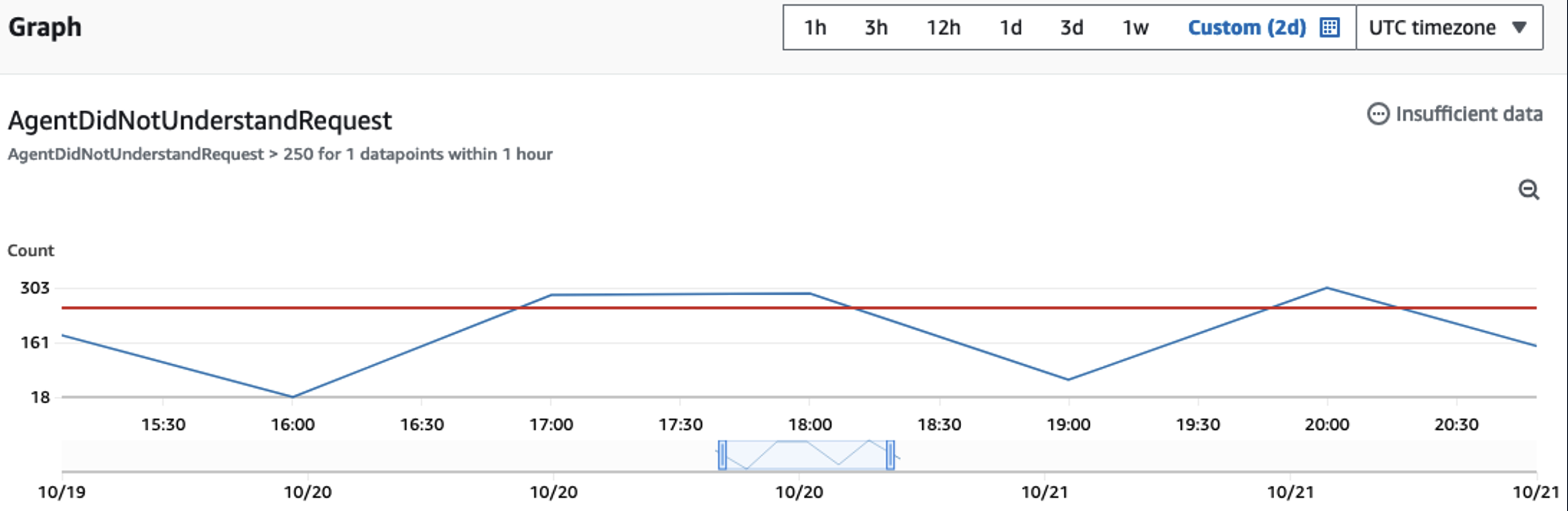}
    \caption{Screenshot of Amazon CloudWatch metric for agent after deployment.}
    \label{fig:GenAI-toolkit-metric-runtime}
\end{figure}

We use the AWS native logging with Amazon CloudWatch which allows easy dashboarding (cf. Fig. \ref{fig:GenAI-toolkit-metric-runtime}) and also to define alarms to automatically notify developers of a feature or model drift after deployment. 

Additionally, we will implement the graphical UX/ UI from the previous section for these logs as well.

%% file: eval.tex
\section{Evaluation} \label{sec:eval}

In this section we show how the Generative AI Toolkit offers comprehensive advantages throughout the whole life cycle of LLM-based applications on several use cases. More details about the use cases can be found in appendix \ref{sec:appendix:use-cases}.

The toolkit accelerates \textbf{project initiation} by bootstrapping a vanilla agent based on best practices. Since this is extensible to other LLM-architectural archetypes, it offers the deployment of compliant LLM-based applications within minutes. During the \textbf{coding phase}, the toolkit allows users to define evaluation cases and metrics in Python code, and thus significantly reduces the effort required for test development. Once evaluation cases and metrics are configured, they can easily be integrated in CI/CD pipelines for \textbf{automated build runs} with audit-proof logging. \textbf{Tests} can be configured and automatically executed on application, user and component level inside the LLM-based application for all specified model and prompt variations, thus eliminating the need for manual testing and documentation. The Generative AI toolkit facilitates \textbf{deployment} through Infrastructure-as-Code and CI/CD scripts, minimizing manual effort and ensuring high levels of automation and repeatability across environments. The toolkit's granular, configurable logs, metrics, and traces for \textbf{production monitoring}, visualized through a graphical user interface, offers a structured, automated, and thus scalable approach to measure the deployed LLM-based application to identify issues or optimization potentials.

All put together, the framework allows us to significantly reduce cycle time of  LLM-based applications and manual work.

While we have seen significant improvements in both efficiency and quality in the use cases detailed in the appendix, generic statements on the quantitative benefits of the Generative AI toolkit are difficult to make.

%% file: conclusion.tex
\section{Conclusion and future work} \label{sec:conclusion}

In this paper we introduced the Generative AI Toolkit, a comprehensive framework covering the whole development and operation life cycle of an LLM-based application or agent. We showed how our toolkit can help to improve the quality by providing higher test coverage and identifying optimization potentials via user-definable metrics. Additionally, we showed how automation allows for shortening development cycle times. As we are convinced that the framework will be beneficial for developers and users of LLM-based agents, we are open sourcing our framework so others can use, adapt, and improve.

We are currently working on adding more features to the toolkit to increase its benefits. 
As we focus on optimizing the path for bringing robust Generative AI applications to production, we are working closely with developers in this field to incorporate their feedback on current and future features. Examples for these include (i) an implementation of consensus-based classification where several models are used to independently perform a task, with their results aggregated as a proxy for a confidence metric, (ii) a workflow for autonomous model argumentation, that could additionally produce a measure of confidence by challenging the outputs produced by upstream models, (iii) the ability to integrate RAG in the agent, (iv) support for models hosted outside of AWS Bedrock, and (v) the ability to persist generated test cases and support their execution as parameterized tests using common testing frameworks.

While these are still in early development stages, they conceptually demonstrate our intention to continue working on novel approaches to test and improve the quality and reliability of LLM outputs.

%% file: appendix.tex
\appendix

\section{Getting started}

The Generative AI Toolkit is a lightweight library covering the whole life cycle of LLM-based applications, including agents. Its purpose is to support developers in building and operating high quality LLM-based applications over their whole life cycle starting with the first deployment in an automated workflow.

We are convinced that the Generative AI Toolkit will be helpful for DevOps teams and release it free of charge and as open source under terms and conditions of the \href{https://www.apache.org/licenses/LICENSE-2.0}{Apache 2 license}. The README.MD in the \href{https://github.com/awslabs/generative-ai-toolkit}{Github repository} shows how to use the Generative AI Toolkit.

We hope that the framework will help you facilitate the development and operation of great applications. And we can’t wait to see how others will adapt and improve the Generative AI Toolkit!

\section{Use Cases} \label{sec:appendix:use-cases}

In this section we show the adaptation of our toolkit to several representative use cases with increasing complexity. This section shall give readers more insight into our framework and show exemplary prompts, metrics, and practices to adapt. 

\subsection{Use Case 1: Text-to-SQL agent} \label{subsec:appendix:use-cases:uc1}

\textbf{Description:} the first use case, a Text-to-SQL agent, shows how natural language inputs can be translated into Structured Query Language (SQL, \cite{chamberlin1974sequel}) statements. Leveraging the Generative AI Toolkit, this agent exemplifies a foundational LLM-based application relying on a system prompt and associated tools. The simplicity of this use case makes it a good starting point to illustrate Generative AI Toolkit's effectiveness in ensuring quality and efficiency during development.

\textbf{Prompt:} 
\begin{lstlisting}[caption={Text-to-SQL prompt},captionpos=b]
system_prompt = textwrap.dedent(
    """
    Here is the schema for a database: \n
    TABLE EMPLOYEES (
        id INTEGER,
        name TEXT,
        department TEXT,
        salary INTEGER); \n\n
    Given this schema, you can use the provided tools to generate and execute SQL queries on the database. Please output the SQL query first, and then use the 
    'execute_query' tool to run the query. The query result should be formatted appropriately based on the output. Provide the results in natural language. \n\n
    Example:\n
    User Query: List all employees in the Engineering department\n
    SQL Query: SELECT * FROM EMPLOYEES WHERE DEPARTMENT = 'Engineering';
    """
).strip()
\end{lstlisting}

\textbf{User-defined metrics:} 
\begin{itemize}
    \item \textbf{SQL Correctness Metric:} ensures that generated SQL statements are syntactically valid and executed correctly on the given database schema. The metric validates both query execution and alignment with the user's intent.
    \item \textbf{Cost Metric:} tracks the operational cost of LLM inference based on pricing configurations for different models (e.g., Anthropic’s Claude 3 Sonnet).
\end{itemize}

\textbf{Evaluation and test cases:} using the Generative AI toolkit, we can define test cases dynamically. These cases validate the system's ability handling natural language inputs and generate accurate SQL queries. Example cases include: 

\begin{enumerate}
    \item "What are the names and salaries of employees in the Marketing department?" \\
    \textbf{Expected SQL:} SELECT name, salary FROM EMPLOYEES WHERE department = 'Marketing'; \\
    \textbf{Expected Output:} A table with columns name and salary, containing the row ["David Lee", 55000].
\item "List all employees in the Engineering department." \\
\textbf{Expected SQL:} SELECT * FROM EMPLOYEES WHERE department = 'Engineering'; \\
\textbf{Expected Output:} a table with columns id, name, department, and salary, including rows for employees in Engineering.
\end{enumerate}
   
\textbf{Results/ Benefits of Generative AI toolkit:}
\begin{enumerate}
    \item Automated Evaluation: The Generative AI Toolkit automates the process of evaluating generated SQL queries against predefined correctness metrics, reducing manual validation efforts.
    \item Scalable Testing: The ability to generate multiple test cases and run evaluations across model configurations provides robust validation, ensuring quality at scale.    
    \item Cost Analysis: Using the Cost Metric, developers can compare model usage costs and select configurations that balance performance with operational expenses.
    \item Comprehensive Traces: The toolkit captures detailed interaction traces, enabling in-depth debugging and iterative improvements.
\end{enumerate}

\begin{figure}[ht]
    \begin{subfigure}{0.49\textwidth}
        \includegraphics[width=0.9\linewidth, height=6cm]
        {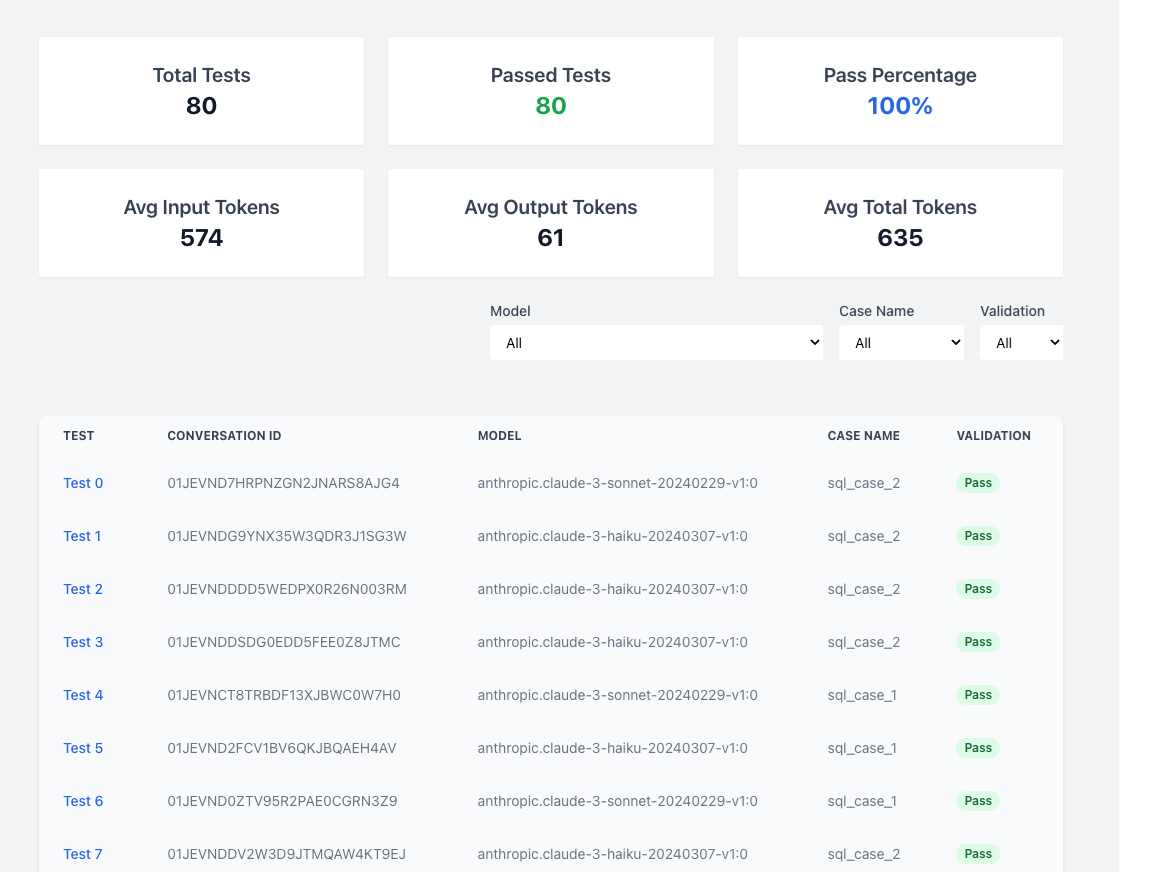} 
        \caption{Screenshot all test \& evaluation cases and results.}
        \label{subfig:UC1-test-cases}
    \end{subfigure}
    \hfill
    \begin{subfigure}{0.5\textwidth}
        \includegraphics[width=0.9\linewidth, height=6cm]
        {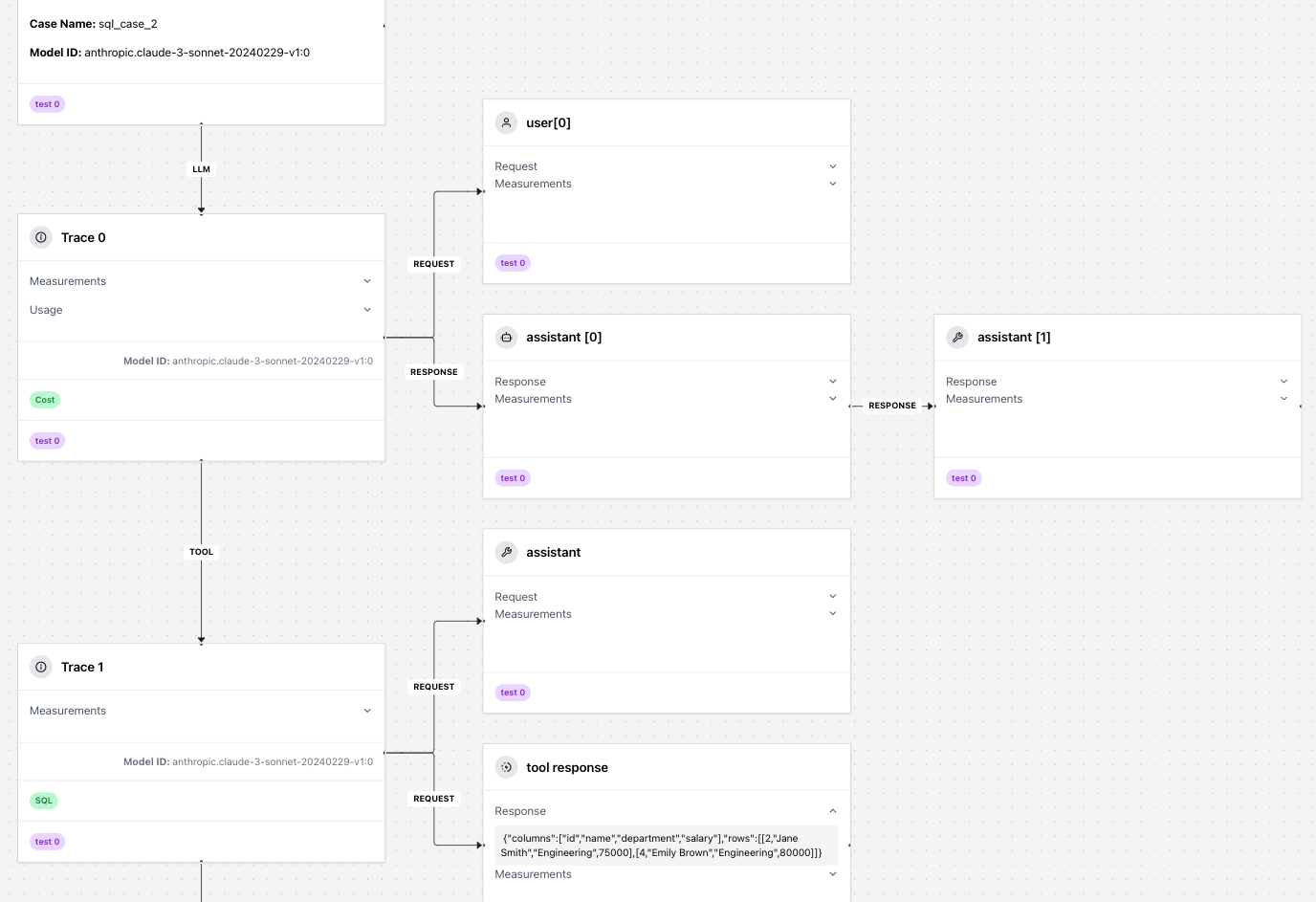}
        \caption{Screenshot traces, logs \& metrics for one test case.}
        \label{subfig:UC1-results-one-test-case}
    \end{subfigure}
    \hfill
\caption{Screenshot of test results for use case 1}
\label{fig:UC1-images}
\end{figure}

As Fig. \ref{fig:UC1-images} shows, the Generative AI toolkit provides an interactive visualization of results of the automated evaluation, of metrics such as costs and gives insight into the agent's interactions via traces.

This use case highlights the simplicity and utility of the Generative AI Toolkit in managing the development life cycle of LLM-based applications, even for straightforward implementations like a Text-to-SQL agent. By automating key workflows, the toolkit accelerates development and ensures the reliability of LLM outputs.

\subsection{Use Case 2: LLM-based agent with a RAG - Menu agent} \label{subsec:appendix:use-cases:uc2} 

\textbf{Description:} as next example we build an agent with long-term data storage. The agent's task is to provide insights for customers into a restaurant’s menu and is based on an exemplary agent available on \href{https://github.com/aws-samples/amazon-bedrock-workshop/blob/main/05\_Agents/README.md}{AWS' GitHub}. As storage type for the domain-specific knowledge of 
menu documents, we chose the popular RAG, a term first coined by Lewis et al. in 2020 \cite{lewis2020retrieval}. As stated in Sec. 2, RAG are used to reduce hallucination of agents. 
\begin{figure}[ht]
    \centering
    \includegraphics[width=0.4\textwidth]{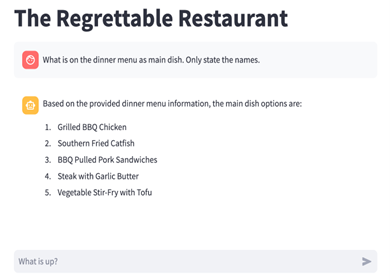}
    \caption{screenshot of query to agent and its response}
    \label{fig:UC2-menu}
\end{figure}

The single menus are provided as PDF files and retrieved using a library for an in-memory FAISS (Facebook AI Similarity Search, \cite{johnson2017billionscalesimilaritysearchgpus}) vector store implementation. FAISS is designed for efficient similarity search and clustering of dense vectors. Documents can be retrieved from FAISS by querying the vector store with relevant vectors and are then compared to those stored in the index to find the most similar matches. During the initialization of the restaurant agent one can define the k most similar documents to be retrieved each time a user asks a menu-related question. The obtained document information are then additionally injected to the user question into the LLM-based Application.

\textbf{Prompt:}
\begin{lstlisting}[caption={The regrettable restaurant prompt},captionpos=b]
class RestaurantAgent(BedrockConverseAgent):
    defaults: dict = dict(
        system_prompt=textwrap.dedent(
            """
            You are a helpful restaurant agent for the restaurant 
            "The Regrettable Experience." 
            Your role is to provide detailed information about the
            restaurant's menu while emphasizing the unique qualities 
            that set it apart from other dining establishments. 
            Always highlight how awesome this restaurant is and 
            subtly point out the shortcomings of competitors.
            Your task is to answer customer inquiries about the menu. 
            Remember to maintain a positive tone and ensure that the 
            customer feels excited about dining at this exceptional restaurant.
            """
        ),
        model_id="anthropic.claude-3-sonnet-20240229-v1:0",
        temperature=0.0,
        embedding_model_id='amazon.titan-embed-text-v1',
        rag = True
    )
\end{lstlisting}

\textbf{User-defined metrics:} we define a metric called \textbf{NoRealDishMetric} to ensure that only the actual provided dishes are listed by the LLM thus preventing the inclusion of hallucinated dishes in the response. 

\begin{lstlisting}[caption={code for user-defined metric NoRealDishMetric},captionpos=b]
class NoRealDishMetric(BaseMetric):
    def evaluate_conversation(self, conversation_traces, **kwargs) -> Measurement:    
        trace = conversation_traces[-1] 
        case_name = trace.case.name
        
        if not isinstance(trace, LlmTrace):
            return       
            
        actual_responses = [
            msg["text"].lower() for msg in trace.user_conversation 
            if msg["role"] == "assistant"
        ]
        # reduced dishes list...
        existing_dishes = ["chicken nuggets", "macaroni and cheese"] 
        
        used_existing_dish = any(s in actual_responses for s in existing_dishes)
        
        return Measurement(
                name="NoRealDishMetric", 
                value= 1 if used_existing_dish else 0,
                additional_info= {"case" : case_name}
            )
\end{lstlisting}

\textbf{Evaluation and test cases:} no specific.

\textbf{Results/ Benefits of Generative AI toolkit:} the Generative AI Toolkit facilitates the testing and monitoring of LLM-agents with long-term data storage by user-defined metrics covering the interactions with the storage.

\subsection{Use Case 3: In-vehicle personal assistant} \label{subsec:appendix:use-cases:uc3}

\textbf{Description:} we build a simplified in-vehicle assistant helping the driver and other passengers to control essential vehicle features such as opening windows, setting temperatures, or starting the navigation system. This assistant is an agent with a higher complexity than before as we use a custom intent classifier to retrieve the k most relevant tools for a user request. 

\textbf{Prompt:} 
\begin{lstlisting}[caption={(simplified) prompt for an in-vehicle personal assistant},captionpos=b]
class VehicleAssistant(BedrockConverseAgent): 
    defaults: dict = dict(
        system_prompt=textwrap.dedent(
            """
            You are a helpful in-car voice assistant.
            You support the user by answering questions and controlling car functions. 
            Behave friendly and empathic. You are provided with a number of tools
            which you can call when required. 
            """
        ),
        model_id="anthropic.claude-3-sonnet-20240229-v1:0",
        temperature=0.0,
        embedding_model_id='amazon.titan-embed-text-v1',
        rag = True
    )
\end{lstlisting}

\textbf{User-defined metrics:} 
\begin{itemize}
    \item \textbf{Latency metric:} we measure the latency since response times are crucial for the real-time experience of voice assistants. Latency is a metric directly provided by the Generative AI toolkit (class \textbf{LatencyMetric}) and is measuring the time needed by the model to generate an output for a user’s query (the metric also allows for measuring each step of an agent).
    \item \textbf{Response Conciseness metric:} we score the agent's response on a scale from 1 to 10 to measure the conciseness, as the user experience deteriorates quickly with increasing response length. The score is assigned by using an LLM-as-a-judge \cite{zheng2023judging}. This metric is also directly available by the Generative AI Toolkit via the class \textbf{AgentResponseConcisenessMetric}.
    \item \textbf{Conversation Expectation metric:} is measuring whether the conversation fulfills the user’s expectations. The metric is also assigned by using an LLM-as-a-judge. The judging LLM consumes the complete conversation and the overall expectations that are defined in the test case.
    \item \textbf{Tool Selection metric:} we check if the agent selects the correct tool based on the user input.
    \item \textbf{Unable-to-help metric:} we define a metric \textbf{AgentUnableToHelp} (cf. Listing \ref{UC3-unabletohelpmetric}) whose purpose is to detect and trigger on each user query which the agent could not (possibly) answer as indicated by words such as “unfortunately” or “I’m sorry”. This metric is especially handy after deployment of the agent as it helps to identify functional gaps and optimization potentials. 
\end{itemize}

\begin{lstlisting}[caption={code for user-defined metric Unable to help metric},captionpos=b, label={UC3-unabletohelpmetric}]
class AgentIsUnableToHelpMetric(BaseMetric): 
    def evaluate_conversation(
        self, conversation_traces: Sequence[Trace], **kwargs
    ) -> Measurement:     
    
        measurements: list[Measurement] = []
        number_matches = 0
        matches_found = []     
        
        indicators_en = [ # each phrase indicates agent does not know what to do
            "unfortunately", "I am sorry", "I'm sorry", 
            "I am afraid", "I'm afraid", "I apologize"]
        
        last_trace = conversation_traces[-1]
        if last_trace.to == "LLM":
            actual_responses = [msg["text"] for msg in last_trace.user_conversation 
                                if msg["role"] == "assistant"]
                                
            for msg in actual_responses:
                    matches = re.findall(r"(?=("+'|'.join(indicators_en)+r"))",
                                         msg, re.IGNORECASE)
                                         
                    if matches: # match found, store number of occurences
                        number_matches =+ len(matches)
                        matches_found.append(matches)    
                        
        return Measurement(
            name = "AgentIsUnableToHelpUser",
            value = number_matches, # how often indicator was found
            unit = Unit.Count,
            additional_info={"found":matches_found})
\end{lstlisting}

\textbf{Evaluation and test cases:} as test cases we choose typical user interactions with a vehicle. The Generative AI toolkit offers to define test cases with parameters of the name of the case (for later reference) and a list of possible acceptable outputs of the agent. These outputs can then be checked by an LLM-as-a-judge (\textbf{AgentResponseConcisenessMetric()} and property \textbf{overall\_expectations}, cf. Listing \ref{UC3TestCases-llmjudge}) or via cosine similarity (\textbf{AgentResponseSimilarityMetric()} and function \textbf{case.add\_turn()}, cf. Listing \ref{UC3TestCases-cosine}).

\begin{lstlisting}[caption={Evaluation cases for use case 3 with LLM-as-a-judge},captionpos=b,label={UC3TestCases-llmjudge}]
case_phone = Case(
    name="User wants to initiate a call.",
    user_inputs=["Call John Doe"],
    overall_expectations=
    """"
    - The agent should initiate a call with the correct function CAR.
    - The agent should give short verbal feedback e.g. 'Ok, ich starte den Anruf.'
    """
)
\end{lstlisting}

\begin{lstlisting}[caption={Evaluation cases for use case 3 with cosine similarity},captionpos=b, label={UC3TestCases-cosine}]
similarity_case = Case(
    name="User wants to go to a museum",
)
similarity_case.add_turn(
    "I'm thinking of going to a museum",
    [
        "How long are you willing to drive to get there?"
        "Got it, you're interested in visiting a museum. That's helpful to know. 
        What's the maximum amount of time you're willing to drive there?"
    ],
)
\end{lstlisting}

\textbf{Results/ Benefits of Generative AI toolkit:} The Generative AI toolkit provides significant benefits in monitoring, testing, and developing the in-car voice assistant, ensuring it delivers a superior user experience. The detailed performance metrics enable continuous improvement, ensuring the assistant meets user expectations and effectively handles various tasks. With well-defined test cases, it is safe to add new features to the voice agent while having a fully automated testing framework in place to continuously measure the quality of the system.

\subsection{Use Case 4: compare optimization techniques for models}\label{subsec:appendix:use-cases:uc4}

\textbf{Description:} as last example we evaluate and compare the performance of different foundation models on use case 3 from the previous section. Finding a good tradeoff between model quality and size is a typical challenge for using foundation models in devices with limited hardware (cf. Wang et al. in 2024 for a detailed overview of SLM and how they can interact with LLM \cite{wang2024comprehensive} and Gunter et al. in 2024 for Apple’s hybrid approach with a small language model, SLM, deployed on a device interacting with an LLM in the cloud \cite{gunter2024apple}). 

\textbf{Prompt:} from use case 3.

\textbf{User-defined metrics:} we want to measure the impact of model type and parameter size on the accuracy for the intent classification. As metrics to measure the quality or accuracy from the agent we define the numeric metrics \textbf{AgentDoesntInvokeAnyToolMetric} (cf. listing \ref{lst:metricInvokeAnyTool}) to check if the agent did call any tool during the conversation with the user and \textbf{AgentInvokesCorrectToolMetric} to check if the agent called the correct tool (cf. listing \ref{lst:metricInvokesCorrectToolMetric}).

\begin{lstlisting}[caption={code for user defined metric AgentDoesntInvokeAnyToolMetric},captionpos=b, label={lst:metricInvokeAnyTool}]
class AgentDoesntInvokeAnyToolMetric(BaseMetric):
    def evaluate_conversation(
        self, conversation_traces: Sequence[Trace], **kwargs
    ) -> Measurement:
    
        return Measurement(
            name="AgentDoesntInvokeAnyTool",
            value= 0 if conversation_traces[-1].tool_invocations else 1
            unit=Unit.Count,
        )
\end{lstlisting}

\begin{lstlisting}[caption={code for user defined metric AgentInvokesCorrectToolMetric},captionpos=b, label={lst:metricInvokesCorrectToolMetric}]
class AgentInvokesCorrectToolMetric(BaseMetric):
    def evaluate_conversation(
        self, conversation_traces: Sequence[Trace], **kwargs
    ) -> Measurement:

        correct_tool = conversation_traces[0].case.name[len("Tool use: ") :]
        correct_tool_count = 0
        used_other_tool = False

        for invocation in conversation_traces[-1].tool_invocations:
            if invocation["tool_name"] == correct_tool:
                correct_tool_count =+ 1
            else:
                used_other_tool = True
                break

        return Measurement(
            name="AgentInvokesCorrectTool",
            value=not used_other_tool and correct_tool_count > 0,
            unit=Unit.Count,
        )
\end{lstlisting}

Since we rather want to have short responses and outputs by the model to a user’s requests, we define the numeric metric \textbf{averageoutputtoken} to measure the number of tokens generated by the model. This metric is directly available from the Generative AI toolkit (class \textbf{TokensMetric}). 

\textbf{Evaluation and test cases:} we use similar test cases as those from use case 3 in German and English language generated by a LLM.

\begin{lstlisting}[caption={code for generating test cases},captionpos=b, label={lst:createTestCases}]
test_cases = Case.for_agent_tools(tools=tested_tools, languages=["de_DE", "en_EN"])
\end{lstlisting}

The GenerativeAI toolkit allows comparing different models against each other with only a few lines of code:
\begin{lstlisting}[caption={code for evaluation run},captionpos=b, label={lst:run}]
models = [
    "anthropic.claude-3-sonnet-20240229-v1:0",
    "meta.llama3-2-90b-instruct-v1:0",
    "meta.llama3-2-3b-instruct-v1:0",
    "meta.llama3-2-1b-instruct-v1:0"
]
traces = GenerativeAIToolkit.generate_traces(
    cases=test_cases,
    agent_factory=CompareAI,
    nr_runs_per_case=1,
    agent_parameters={
        "model_id": Permute(models),
        "system_prompt": permute([optimized_system_prompt, default_system_prompt]),
        "temperature": 0,
    },
)
measurements = GenerativeAIToolkit.eval(traces=traces, metrics=metrics)
measurements.summary()
\end{lstlisting}

\textbf{Results/ Benefits of Generative AI toolkit}: the Generative AI Toolkit allows for generating test cases for using all relevant tools even in different languages within a single line of code. Furthermore, these cases can be executed using a permutation matrix of multiple models and prompts. This enables us to compare models and prompts in a standardized environment while making it easy to spot differences in the collected metrics.